\documentclass{article}

\usepackage{microtype}
\usepackage{graphicx}
\usepackage{booktabs}
\usepackage{longtable}
\usepackage{tikz}
\usetikzlibrary{arrows.meta,positioning,calc}
\PassOptionsToPackage{hyphens}{url}
\usepackage{hyperref}
\usepackage{xcolor}

\usepackage[preprint]{icml2026}

\icmltitlerunning{Lacuna: A Research Map for Machine Learning}

\begin{document}

\twocolumn[
  \icmltitle{Lacuna: A Research Map for Machine Learning}

  \begin{icmlauthorlist}
    \icmlauthor{Martin Weiss}{tasc,mila,poly}
    \icmlauthor{Miles Q.~Li}{mcgill,tasc}
    \icmlauthor{Alejandro H.~Artiles}{tasc}
    \icmlauthor{Yacine Mkhinini}{tasc}
    \icmlauthor{Chris Pal}{mila,poly}
    \icmlauthor{Hugo Larochelle}{udem,mila}
    \icmlauthor{Nasim Rahaman}{tasc}
  \end{icmlauthorlist}

  \icmlaffiliation{tasc}{Tiptree Advanced Systems Corporation}
  \icmlaffiliation{mila}{Mila}
  \icmlaffiliation{poly}{Polytechnique Montreal}
  \icmlaffiliation{mcgill}{McGill University}
  \icmlaffiliation{udem}{Universite de Montreal}
  \icmlcorrespondingauthor{Martin Weiss}{martin@tiptreesystems.com}
  \icmlkeywords{research maps, knowledge graphs, problem formulation, literature discovery, machine learning}

  \vskip 0.3in
]

\printAffiliationsAndNotice{}

\begin{abstract}
Lacuna is a research map for machine learning that uses LLMs to turn papers and scholarly metadata into markdown summaries, concept elements, research directions, and research proposals. Each item keeps links to the primary source records and papers that support it.
We release the map with web, markdown, and MCP interfaces. 
Across LitSearch, Multi-XScience-CS/ML, and ScholarQA-CS-ML, Lacuna outperforms OpenScholar with the strongest gains on LitSearch retrieval (Recall@10 0.538 vs.\ 0.424 for OpenScholar v3).
We also evaluate Lacuna Deep Research, a multi-stage report agent over the map, on 25 ReportBench-ML survey tasks: Lacuna Deep Research reaches 0.052 citation F1, 0.339 citation precision, 99 expert-reference hits, and 7.82/10 RACE report quality, while GPT-Researcher reaches 0.039 F1, 0.290 precision, 72 hits, and 5.24/10 RACE.
\end{abstract}

\section{Introduction}
\label{sec:intro}

Lacuna is a large-scale research map for machine learning (ML): a served, linked, paper-grounded layer over the literature that humans and agents can browse, search, cite, and build on. Instead of treating scientific papers as isolated PDFs that must be reread for every query, Lacuna turns scholarly records and paper content into inspectable objects: paper summaries, concept elements, research directions, and generated research proposals.

The goal is to make the intermediate structure of research reusable. A user or LLM agent should be able to start from a seed topic, a broad survey request, or a concrete literature-review question and immediately enter a paper-grounded neighborhood of methods, limitations, evidence, and candidate hypotheses. Lacuna therefore acts as research infrastructure rather than only as a retrieval system: it gives models and researchers a map to navigate before they write an answer, report, or proposal.

This map is useful for several tasks. In \emph{research problem formulation}, Lacuna helps move from an underspecified topic, for example ``automated theorem proving with proof assistants'', to a scoped research question with supporting observations and limitations. In \emph{deep research}, Lacuna provides the evidence substrate for multi-stage report agents that need broad coverage, parallel exploration, and cited synthesis. In literature-review and scientific-QA settings, Lacuna supplies compact evidence that can be compared against established literature-search and synthesis baselines.

\begin{figure*}[t]
\centering
\includegraphics[width=0.91\textwidth,height=0.255\textheight,keepaspectratio]{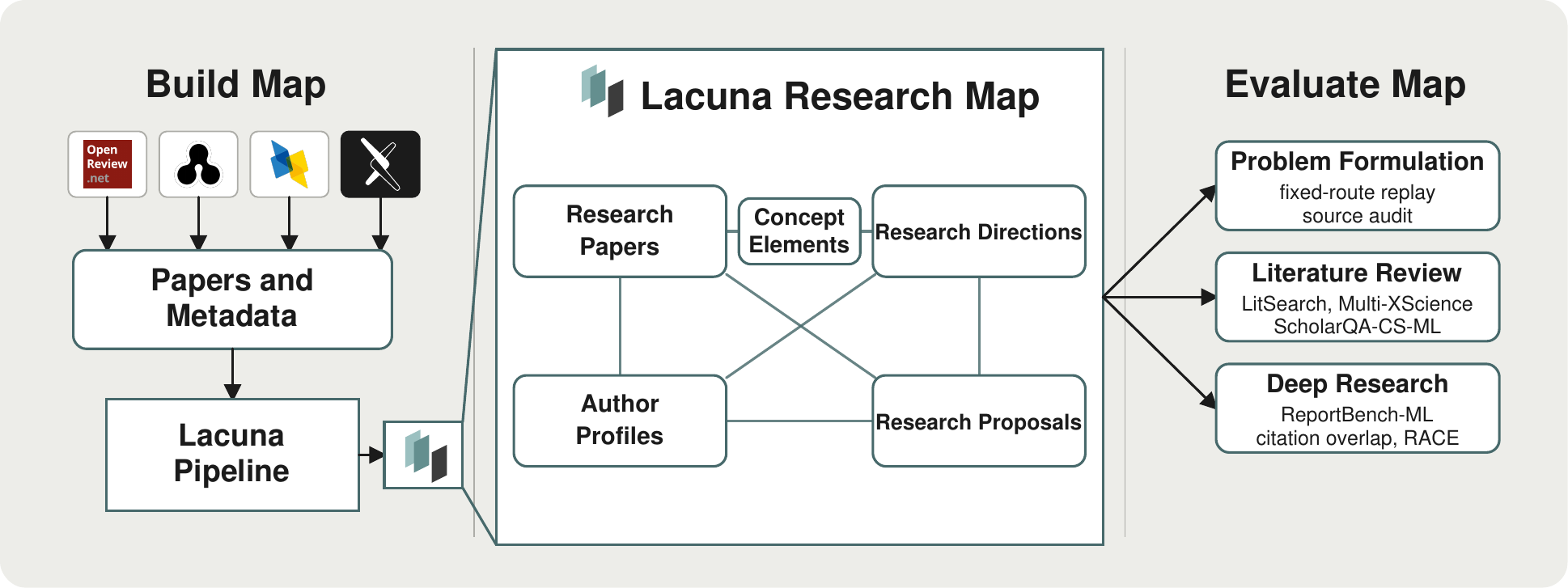}
\vspace{0.04in}
\caption{Overview of Lacuna. Scholarly records from OpenReview, OpenAlex, DBLP, and arXiv flow through the Lacuna pipeline into a linked research map of research papers, concept elements, research directions, author profiles, and research proposals. The map is evaluated in three settings: problem formulation with fixed-route replay and source audit, literature review with LitSearch, Multi-XScience-CS/ML, and ScholarQA-CS-ML, and deep research with ReportBench-ML citation overlap and RACE; results are reported in Sections~\ref{sec:problem-formulation-framework}--\ref{sec:deep-research}.}
\label{fig:lacuna-overview}
\end{figure*}

Lacuna hosts this map\footnote{\url{https://lacuna.tiptreesystems.com}.} over 733{,}795 catalogued papers. It exposes web pages, markdown pages at \texttt{/md}, a site-wide schema, and an MCP interface. A model can fetch compact pages, follow typed links, compare neighboring directions, and end at a report, answer, or research proposal whose claims can be audited against source-linked evidence. Figure~\ref{fig:lacuna-overview} summarizes the build-map-evaluate flow; Figure~\ref{fig:research-map-scale} shows the layered map and its live scale.

\textbf{Contributions.}
\vspace{-0.55em}
\begin{enumerate}
\setlength{\itemsep}{0.15em}
\setlength{\topsep}{0pt}
\setlength{\parsep}{0pt}
    \item \emph{Lacuna}, a served research map for ML with 733{,}795 paper pages, 15{,}259{,}720 concept elements, 27{,}017 research directions, and 38{,}000 research proposals.
    \item A provenance-preserving build pipeline from scholarly records to paper summaries, concept elements, directions, and proposals.
    \item Human- and agent-facing interfaces: web pages, markdown variants, \texttt{/llms.txt}, and MCP access.
    \item Evaluations on problem formulation, OpenScholar literature-review benchmarks, and ReportBench ML deep research against GPT-Researcher, STORM, and LangChain ODR.
\end{enumerate}

\section{Related Work}
\label{sec:related}

\textbf{Scholarly corpora and scholarly knowledge graphs.} Open scholarly infrastructure such as OpenAlex~\citep{priem2022openalex} and S2ORC~\citep{lo2020s2orc} provides large-scale metadata, identifiers, and machine-readable paper text. Document representation work such as SPECTER~\citep{cohan2020specter} learns citation-informed embeddings for scholarly search and recommendation, while scholarly knowledge graph systems such as the Open Research Knowledge Graph~\citep{jaradeh2019orkg} represent scientific contributions in machine-actionable form. Lacuna builds on this infrastructure with generated, paper-grounded entries for problem formulation.

\textbf{Information foraging and exploratory search.} Information foraging theory models information seeking as a tradeoff between information value and the costs of locating and consuming it~\citep{pirolli1999information}. Its notions of information patches and information scent fit scholarly navigation: researchers sample local cues, enter promising topic regions, and abandon weak paths as expected value changes. Lacuna uses this view by making research directions navigable topic regions with low-cost previews and grounded relationships.

\textbf{Scientific QA and retrieval-augmented synthesis.} PaperQA~\citep{lala2023paperqa} and OpenScholar~\citep{asai2026openscholar} answer well-formed scientific questions by retrieving evidence and generating citation-backed responses; SciFact verifies claims against papers~\citep{wadden2020fact}. Our workflow targets the upstream stage of \emph{forming} the question, before retrieval systems generate answers. Section~\ref{sec:scholarqa} evaluates whether Lacuna's compact evidence can also support citation-backed synthesis on ScholarQA-CS-ML.

\textbf{Literature-based discovery.} The notion of research problem formulation targeted here overlaps with literature-based discovery, which studies methods for finding implicit hypotheses in scientific text~\citep{thilakaratne2019systematic}. Lacuna's research proposal pages are produced using Alien Science sampling~\citep{artiles2026aliensciencesamplingcoherent}; we treat them as hypotheses for researcher review.

\section{The Lacuna Research Map}
\label{sec:map}

\begin{figure*}[t]
\centering
\includegraphics[width=0.94\textwidth,height=0.36\textheight,keepaspectratio]{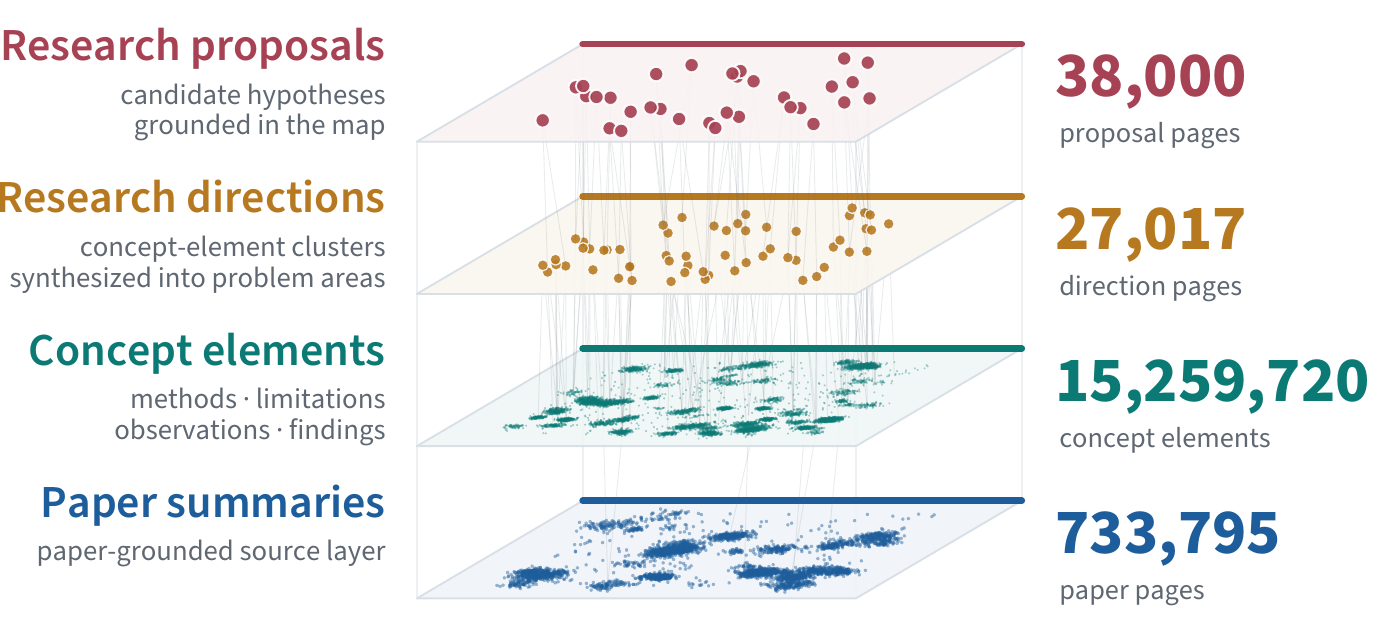}
\vspace{0.04in}
\caption{Live scale of the primary generated layers in the Lacuna research map. Paper summaries form the source layer; extracted concept elements are embedded and clustered into research directions; research proposals are generated from directions and supporting papers. Right-side labels report live cardinalities: 733{,}795 paper pages, 15{,}259{,}720 concept elements, 27{,}017 direction pages, and 38{,}000 proposal pages.}
\label{fig:research-map-scale}
\end{figure*}

The map has four primary generated layers: paper summaries, concept elements, research directions, and research proposals. Figure~\ref{fig:research-map-scale} shows these layers and their live cardinalities.

\textbf{Lacuna Contents.} Lacuna contains several generated page and record types. A \emph{core-idea paper summary} is a markdown document generated from the paper text and metadata to capture its main technical ideas. A \emph{paper summary with figures} is a compact paper summary that combines text with extracted and selected figures. A \emph{concept element} is a one- or two-sentence statement of a core idea, method, limitation, or empirical observation extracted from a core-idea paper summary. Concept elements are embedded, linked to papers, and clustered. A \emph{research direction} is an approximately two-page synthesis of a cluster of concept elements into a recurring problem, method family, or opportunity area. A \emph{research proposal} is a generated hypothesis sampled from research directions and supporting papers. The production map used \texttt{gemini-flash-3.1-preview} for summaries, concepts, and directions, and \texttt{gemini-3.1-pro-preview} for proposals~\citep{geminiteam2026gemini31}.

\textbf{Lacuna Interface.} Lacuna exposes four interfaces: RRF (reciprocal-rank-fusion) search; \texttt{/md} pages with 5--30\,KB of cleaned markdown; typed links among map pages; and an MCP artifact documenting the page schema for LLM tools. Section~\ref{sec:construction-main} and Appendix~\ref{app:construction} describe construction.

\section{The Lacuna Build Pipeline}
\label{sec:construction-main}

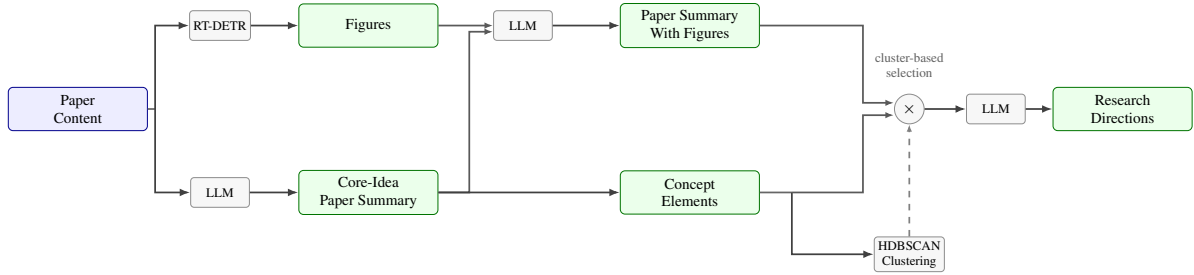
\begin{figure*}[t]
\centering
\resizebox{0.92\textwidth}{!}{\begin{tikzpicture}[
  font=\scriptsize,
  source/.style={
    draw,
    rounded corners=2.2pt,
    align=center,
    minimum width=2.25cm,
    minimum height=0.70cm,
    inner sep=3pt,
    fill=blue!7,
    draw=blue!55!black
  },
  generated/.style={source, fill=green!8, draw=green!45!black},
  process/.style={
    draw,
    rounded corners=1.6pt,
    align=center,
    minimum width=0.95cm,
    minimum height=0.48cm,
    inner sep=1.8pt,
    font=\tiny,
    fill=black!3,
    draw=black!45
  },
  selector/.style={
    draw,
    circle,
    align=center,
    minimum size=0.48cm,
    inner sep=0pt,
    font=\scriptsize,
    fill=black!3,
    draw=black!45
  },
  note/.style={font=\tiny, text=black!62, align=center},
  flow/.style={-{Latex[length=1.7mm]}, thick, black!75},
  stem/.style={thick, black!60},
  mergeflow/.style={-{Latex[length=1.4mm]}, thick, black!65},
  controlflow/.style={-{Latex[length=1.2mm]}, dashed, thick, black!50}
]
\path[use as bounding box] (-7.2,-2.85) rectangle (12.0,2.85);

\node[source] (paper-content) at (-6.00,0.00) {Paper\\Content};

\node[process] (rt-detr) at (-3.70,1.35) {RT-DETR};
\node[process] (llm-summary) at (-3.70,-1.35) {LLM};

\node[generated] (figures) at (-1.30,1.35) {Figures};
\node[generated] (paper-summary) at (-1.30,-1.35) {Core-Idea\\Paper Summary};

\node[process] (llm-merge) at (1.20,1.35) {LLM};

\node[generated] (summary-with-figures) at (3.90,1.35) {Paper Summary\\With Figures};
\node[generated] (concept-elements) at (3.90,-1.35) {Concept\\Elements};
\node[process] (hdbscan) at (7.45,-2.35) {HDBSCAN\\Clustering};
\node[selector] (direction-select) at (7.45,0.00) {$\times$};
\node[note, above=0.14cm of direction-select] {cluster-based\\selection};
\node[process] (direction-llm) at (8.85,0.00) {LLM};
\node[generated] (research-directions) at (10.90,0.00) {Research\\Directions};

\coordinate (content-split) at (-4.75,0.00);
\draw[stem] (paper-content.east) -- (content-split);
\draw[flow] (content-split) |- (rt-detr.west);
\draw[flow] (rt-detr.east) -- (figures.west);
\draw[flow] (content-split) |- (llm-summary.west);
\draw[flow] (llm-summary.east) -- (paper-summary.west);
\path (llm-merge.west) coordinate (merge-in-figures);
\path (llm-merge.west) ++(0,-0.10) coordinate (merge-in-summary);
\draw[mergeflow] (figures.east) -- (merge-in-figures);
\draw[mergeflow] (paper-summary.east) -- ++(0.50,0) |- (merge-in-summary);
\draw[flow] (llm-merge.east) -- (summary-with-figures.west);
\draw[flow] (paper-summary.east) -- (concept-elements.west);
\coordinate (direction-concept-split) at (5.55,-1.35);
\coordinate (selection-elbow-summary) at (6.70,1.35);
\coordinate (selection-elbow-concepts) at (6.70,-1.35);
\path (direction-select.west) ++(0,0.10) coordinate (select-in-summary);
\path (direction-select.west) ++(0,-0.10) coordinate (select-in-concepts);
\draw[stem] (concept-elements.east) -- (direction-concept-split);
\draw[flow] (direction-concept-split) |- (hdbscan.west);
\draw[mergeflow] (summary-with-figures.east) -- (selection-elbow-summary) |- (select-in-summary);
\draw[mergeflow] (direction-concept-split) -- (selection-elbow-concepts) |- (select-in-concepts);
\draw[controlflow] (hdbscan.north) -- (direction-select.south);
\draw[flow] (direction-select.east) -- (direction-llm.west);
\draw[flow] (direction-llm.east) -- (research-directions.west);
\end{tikzpicture}}
\caption{Core paper-to-direction generation in Lacuna. Paper content is processed into figures and core-idea summaries; summaries and figures are merged into paper summaries with figures, core summaries produce concept elements, and cluster-based selection supplies summaries and concepts to an LLM synthesis stage that writes research directions.}
\label{fig:pipeline}
\end{figure*}

Lacuna is built by a pipeline. It harvests and reconciles records from OpenAlex, OpenReview, DBLP, arXiv, and venue open-access pages, then attaches paper files or URL replicas to stable paper and metadata records. This stage establishes the identifiers that later generated content refers to. Figure~\ref{fig:pipeline} shows the core paper-to-direction generation subpipeline; Appendix~\ref{app:construction} gives the full stage table.

Within the catalog stage, Lacuna treats author bibliography construction as an authorship-anchored expansion problem. We use OpenReview as the primary identity anchor because its author-maintained records provide a cleaner signal than external author identifiers alone, which can conflate researchers with shared names. For each OpenReview-profiled author, the pipeline harvests that author's OpenReview papers, canonicalizes aliases, and builds an OpenReview-only co-author neighborhood. External sources such as DBLP and OpenAlex are used as recall sources, not as authorities. Papers obtained through an author's external identifiers are retained for that author's bibliography only when they either reconcile to an existing OpenReview paper or when their author lists overlap the author's trusted OpenReview co-author neighborhood. This reduces the risk that external-source disambiguation errors propagate into paper pages, research-direction memberships, and proposal context.

LLM calls then turn papers and metadata into reusable summaries, concepts, directions, and proposals. From paper text and metadata, the pipeline generates core-idea paper summaries; from figures extracted with RT-DETR~\citep{zhao2024detrs} and selected for paper context, it generates paper summaries with figures. Concept-element extraction converts these summaries into one- or two-sentence units that capture methods, observations, limitations, and empirical findings.

The graph-building stages embed concept elements, cluster them with HDBSCAN~\citep{mcinnes2017hdbscan} into research directions, and generate research proposal pages from research directions and supporting papers. Serving stages then build search indexes, relation tables, HTML pages, markdown variants, and MCP access. Each generated item keeps links to the papers, metadata records, concept elements, and research directions that produced it, which lets a researcher inspect evidence from a research proposal back to the underlying paper pages.

\section{Applications and Evaluation}
\label{sec:applications}
\label{sec:evaluation}

We evaluate Lacuna through three application settings: research problem formulation, literature-review question answering, and deep-research report synthesis. The first tests map navigation from a seed idea to a proposal; the second tests compact evidence for search and synthesis; the third tests a multi-stage report agent built on the map.

\subsection{Research Problem Formulation}
\label{sec:problem-formulation-framework}
\label{sec:route}

We target \textbf{research problem formulation}: given a seed query, produce a scoped research question, a small number of supporting observations, and a limitation statement in the form a researcher could circulate to a collaborator. Lacuna supports formulation by precomputing research directions and research proposals; the LLM's role is to choose, connect, and audit these for the researcher.

We instantiate the run on the seed idea \emph{automated theorem proving with proof assistants}. The run uses Lacuna's hosted markdown pages, which can be reached through public routes under \url{https://lacuna.tiptreesystems.com/md} or through direct HTTPS \texttt{GET} requests.

The run visits two research direction pages, four paper pages, and one research proposal page. It starts at a research direction about the autoformalization gap between informal mathematical reasoning and proof-assistant verification, follows a neighboring research direction about iterative compiler feedback, and then inspects paper pages for ProofNet~\citep{azerbayev2023proofnet}, FIMO~\citep{liu2023fimo}, StepFun-Prover~\citep{shang2025stepfun}, and VERINA~\citep{ye2025verina}. These papers provide quantitative anchors: low Codex success on ProofNet, GPT-4's 0\% proof success on FIMO, StepFun-Prover's 70.0\% pass@1 on miniF2F, and VERINA's gap between code generation and proof generation. The run ends at a research proposal page that turns this local neighborhood into a hypothesis about compiler-feedback granularity.

The synthesized final question is: \emph{at what granularity of compiler feedback --- tactic-block, subgoal, or token-level --- does iterative LLM-REPL interaction most efficiently separate fixable translation errors from fundamental deductive gaps?} The run therefore turns the broad seed idea into a concrete, inspectable problem formulation: a specific independent variable, a mechanism to test, and a limitation of current one-shot and iterative proof-generation methods.

\paragraph{Grounding, time, and cost.}
\label{sec:grounding}

We audit the claims used by the run against the markdown pages it cites, marking each as supported by a cited paper page, a research proposal hypothesis, a research direction limitation, or unsupported. The audit covers five claims: two are supported by cited paper pages, one is a research direction limitation, one is a research proposal hypothesis correctly treated as a proposed mechanism rather than a reported result, and one is a deliberately included false claim --- that the route is perfectly clean --- that Lacuna's paper pages and relation tabs let a reader catch and correct. The full per-claim audit with evidence quotes is in Appendix~\ref{app:audit}.

We also compare the time and cost of consuming Lacuna pages with using raw PDFs for the same theorem-proving task. The sequential-PDF baseline has no access to Lacuna; it downloads four papers, extracts their first ten pages with \texttt{pdftotext}, writes a 650--900 word note per paper, and only then synthesizes the final question. Table~\ref{tab:cost} reports the resulting cost and tool use. This fixed run shows work that Lacuna avoids at inference time: paper notes and map context are served as existing content instead of being regenerated inside the run.

\begin{table}[t]
\caption{Time, tool use, and cost for the fixed theorem-proving run; Lacuna serves map pages, while the baseline reads raw PDFs.}
\label{tab:cost}
\centering
\small
\begin{tabular}{lrr}
\toprule
 & Lacuna \texttt{/md} & PDF baseline \\
\midrule
Wall time (s) & 85.5 & 289.2 \\
Claude turns & 8 & 28 \\
Tool calls & 7 & 27 \\
Tool text read & 41.5\,KB & 167.3\,KB \\
Output tokens & 4{,}919 & 14{,}135 \\
Blocking model calls & 1 & 6 \\
Cost (USD) & \$0.171 & \$0.695 \\
\bottomrule
\end{tabular}
\end{table}

\subsection{Literature Review and Grounded Research Question Answering}
\label{sec:grounded-research-qa}
\label{sec:scholarqa}

Lacuna can also be used as evidence infrastructure for answering well-formed ML/AI literature questions. This application is closer to retrieval-augmented question answering than to problem formulation: the user already has a question, and the system must retrieve relevant papers, synthesize an answer, and preserve enough citations or evidence snippets for the answer to be checked.

Here Lacuna acts as the evidence layer rather than the answer generator itself. Its paper pages, concept elements, and citation-linked snippets provide compact context that an answer model can use and that readers can inspect afterward.

The evaluation asks whether Lacuna's research map is useful for literature search and synthesis beyond the fixed problem-formulation run. Table~\ref{tab:benchmark-summary} summarizes the current benchmark set. LitSearch~\citep{ajith2024litsearch} measures paper retrieval for natural-language paper descriptions; Multi-XScience-CS/ML~\citep{lu2020multixscience} measures related-work synthesis on a CS/ML-filtered slice of Multi-XScience; and ScholarQA-CS-ML measures literature QA on a 42-question ML/AI slice of ScholarQA-CS. We compare against OpenScholar v3 where available~\citep{asai2026openscholar}.

\begin{table}[t]
\caption{Literature search and synthesis benchmark summary. Higher is better.}
\label{tab:benchmark-summary}
\centering
\scriptsize
\begin{tabular}{@{}llcc@{}}
\toprule
Benchmark & Metric & OpenScholar & Lacuna \\
\midrule
LitSearch & MRR & 0.3081 & \textbf{0.3585} \\
LitSearch & R@10 & 0.424 & \textbf{0.538} \\
Multi-XScience-CS/ML & score / 5 & 3.467 & \textbf{4.167} \\
ScholarQA-CS-ML & rubric avg. & 0.672 & \textbf{0.694} \\
\bottomrule
\end{tabular}
\end{table}

The results show different strengths of the map. On LitSearch, Lacuna reaches R@10 0.538 and MRR 0.3585 on the full 597-query split, above OpenScholar v3. On Multi-XScience-CS/ML, Lacuna scores 4.167, above OpenScholar and close to original cited abstracts despite partial citation coverage. On ScholarQA-CS-ML, Lacuna-GPT-4o scores 0.694 under the ScholarQABench rubric judge, compared with 0.672 for the released OpenScholar-GPT-4o answer file.

ScholarQA-CS is the 100-question computer-science literature-synthesis subset of ScholarQABench, introduced with OpenScholar in \emph{Nature}~\citep{asai2026openscholar}; its questions were written by PhD-level domain experts and scored against detailed literature-grounded rubrics. We classified the full 100-prompt set and evaluated every prompt marked ML/AI: 42 prompts, or 42\% of ScholarQA-CS, with no score-based filtering. Appendix~\ref{app:scholarqa} gives the subset-selection details and lists all questions in ScholarQA-CS-ML.

\begin{table}[t]
\caption{ScholarQA-CS-ML literature-synthesis evaluation on 42 ML/AI prompts. Scores are ScholarQABench rubric averages; OpenScholar-GPT-4o is released answer + references, and Lacuna-GPT-4o is answer + cited evidence.}
\label{tab:scholarqa}
\centering
\small
\begin{tabular}{@{}p{0.42\columnwidth}cc@{}}
\toprule
Metric & OpenScholar & Lacuna \\
\midrule
Overall & 0.672 & \textbf{0.694} \\
Rubric items & 0.607 & \textbf{0.638} \\
Expertise & 0.826 & \textbf{0.879} \\
Citation coverage & \textbf{0.956} & 0.924 \\
Excerpts & 0.747 & \textbf{0.826} \\
Must-have items & 0.577 & \textbf{0.615} \\
Nice-to-have items & 0.455 & \textbf{0.469} \\
\bottomrule
\end{tabular}
\end{table}

Table~\ref{tab:scholarqa} reports the same-question comparison: Lacuna-GPT-4o scores 0.694, compared with 0.672 for the released OpenScholar-GPT-4o answer file. The gain is concentrated in content and evidence quality: Lacuna-GPT-4o is higher on rubric-item coverage, judged expertise, excerpt availability, and both must-have and nice-to-have rubric items. OpenScholar-GPT-4o remains higher on structural citation coverage, which measures whether claims carry inline bracket citations; it does not measure source entailment.

\subsection{Deep Research Report Synthesis}
\label{sec:deep-research-framework}
\label{sec:deep-research}

Deep research report synthesis is the application of using Lacuna to turn broad ML/AI literature requests into cited survey-style reports. It is intended for requests that require breadth rather than quick single-route lookup: broad overviews, literature reviews, research landscapes, roadmaps, and state-of-the-art comparisons.

Lacuna Deep Research is our instantiation of this application on top of the Lacuna map. It uses the same Lacuna search and page-reading tools as the single-route research agent, but changes the orchestration from one long ReAct trajectory to a breadth-first multi-stage workflow.

The workflow has five stages. First, a constraint extractor turns the user request into structured fields such as topic, cutoff dates, must-cover topics, venues, requested sections, and other freeform constraints. Second, a seed search queries Lacuna research directions with the extracted topic, falling back to must-cover terms when needed, and a planner writes a section outline plus independent research lenses. Third, each lens is dispatched to a separate Lacuna research worker with its own context window, and workers run in parallel over the map. Fourth, a writer synthesizes the lens notes and deduplicated figure attachments into a structured cited report that follows the outline and constraints. Fifth, when enforceable constraints were present, an audit-and-revise pass checks the draft against the original request and revises it to cover required topics and respect date constraints.

The breadth comes from several disciplined workers rather than from loosening a single worker's prompt. The agent runs six high-effort lens workers. Worker failures are isolated: synthesis proceeds from surviving lens notes, and the whole report fails only if every lens worker fails. The agent therefore turns the static research map into an application-level deep-research system while keeping evidence collection grounded in Lacuna pages.

For evaluation, we run Lacuna Deep Research on ReportBench-ML: 25 core-ML survey tasks derived from peer-reviewed survey papers, with cited-paper overlap against each expert survey's reference list as an objective grounding metric~\citep{li2025reportbench}. We compare against three open deep-research report agents:

\begin{itemize}
    \item \textbf{GPT-Researcher}~\citep{gptresearcher2023} is an open-source autonomous research agent that plans web or local-document research, gathers sources, and writes citation-backed reports.
    \item \textbf{STORM}~\citep{shao2024storm} is a Stanford knowledge-curation system that generates long-form, cited articles by discovering perspectives, simulating source-grounded question asking, and writing from the resulting outline.
    \item \textbf{LangChain Open Deep Research}~\citep{langchain2025opendeepresearch} is an open-source LangGraph report agent that scopes a research brief, delegates research to sub-agents, and writes a final report from their findings.
\end{itemize}

Because expert surveys cite roughly 140 papers while deep-research agents cite about 6--16 papers, recall is small for every system; precision, F1, and total expert-reference hits are the most informative citation-overlap metrics. We also report RACE, the DeepResearch Bench report-quality rubric, which scores comprehensiveness, insight/depth, instruction-following, and readability with task-adaptive weights~\citep{du2025deepresearchbench}.

\begin{table}[t]
\caption{ReportBench-ML citation-overlap evaluation. Scores use ReportBench title-overlap matching against expert survey references.}
\label{tab:deep-research}
\centering
\small
\begin{tabular}{@{}lcccc@{}}
\toprule
System & F1 & Prec. & Rec. & Hits \\
\midrule
GPT-Researcher & 0.039 & 0.290 & 0.022 & 72 \\
STORM & 0.015 & 0.250 & 0.008 & 21 \\
LangChain ODR & 0.007 & 0.055 & 0.004 & 13 \\
Lacuna Deep Research & \textbf{0.052} & \textbf{0.339} & \textbf{0.028} & \textbf{99} \\
\bottomrule
\end{tabular}
\end{table}

\begin{table}[t]
\caption{RACE rubric breakdown for ReportBench-ML report quality. Scores are out of 10; Comp. is comprehensiveness, Instr. is instruction-following, and Read. is readability.}
\label{tab:race-breakdown}
\centering
\scriptsize
\begin{tabular}{@{}lccccc@{}}
\toprule
System & Overall & Comp. & Insight & Instr. & Read. \\
\midrule
GPT-Researcher & 5.24 & 5.31 & 4.87 & 4.41 & 7.62 \\
STORM & 2.90 & 3.45 & 2.67 & 1.66 & 4.20 \\
LangChain ODR & 7.42 & 7.48 & 7.23 & 7.22 & 8.16 \\
Lacuna Deep Research & \textbf{7.82} & \textbf{8.01} & \textbf{7.61} & \textbf{7.57} & \textbf{8.34} \\
\bottomrule
\end{tabular}
\end{table}

Table~\ref{tab:deep-research} shows that Lacuna Deep Research is the strongest system on citation overlap in this ML subset. Compared with GPT-Researcher, it has higher citation precision (0.339 vs.\ 0.290), higher citation F1 (0.052 vs.\ 0.039), and more expert-reference hits (99 vs.\ 72). Table~\ref{tab:race-breakdown} shows that Lacuna Deep Research also has the highest judged report quality (7.82 vs.\ 5.24 for GPT-Researcher), with broad gains across comprehensiveness, insight, instruction-following, and readability. All 25 ReportBench ML reports completed and were evaluated.

\section{Conclusion}
\label{sec:conclusion}

Lacuna shows how a served research map can support machine learning research discovery. Its linked summaries, concepts, directions, and proposals let researchers and agents skim, branch, and learn their way through a vast corpus of literature to formulate research problems, answer literature questions, and synthesize survey-style reports.

One natural downstream evaluation is AI-assisted peer review. ReviewerToo identifies novelty and significance assessment as a difficult part of LLM-assisted reviewing, even when models are useful for fact-checking and literature coverage~\citep{sahu2025reviewertoo}; focus-level evaluations similarly find that LLM reviews under-attend to novelty when criticizing papers~\citep{shin2025blindspots}. Lacuna offers a concrete intervention for this setting: provide a review agent with source-linked neighboring papers, research directions, and author-profile context, then measure whether the resulting reviews improve on novelty, significance, and literature-coverage rubrics.

Several limitations remain. Lacuna is a batch-built map, not an online updating system: new papers require rerunning the relevant pipeline stages, and incremental updates are future work. The present map targets machine learning literature; extending it to the rest of computer science or to other scientific fields will require additional source coverage and field-specific quality checks. Lacuna also shifts cost into the initial build. Harvesting records, extracting figures, generating summaries, embedding concepts, clustering directions, and building indexes are expensive upfront; the payoff is lower cost for many later page fetches and navigation runs.

\section*{Acknowledgements}

Martin Weiss acknowledges support from IVADO through its postdoctoral entrepreneur program. Lacuna builds on public scholarly infrastructure, including OpenReview, OpenAlex, DBLP, and arXiv; we thank the maintainers of these services for making large-scale research discovery systems possible.

\bibliography{references}

\begin{thebibliography}{26}
\providecommand{\natexlab}[1]{#1}
\providecommand{\url}[1]{\texttt{#1}}
\expandafter\ifx\csname urlstyle\endcsname\relax
  \providecommand{\doi}[1]{doi: #1}\else
  \providecommand{\doi}{doi: \begingroup \urlstyle{rm}\Url}\fi

\bibitem[Ajith et~al.(2024)Ajith, Xia, Chevalier, Goyal, Chen, and
  Gao]{ajith2024litsearch}
Ajith, A., Xia, M., Chevalier, A., Goyal, T., Chen, D., and Gao, T.
\newblock {LitSearch}: A retrieval benchmark for scientific literature search.
\newblock In \emph{Proceedings of the 2024 Conference on Empirical Methods in
  Natural Language Processing (EMNLP)}, 2024.
\newblock URL \url{https://arxiv.org/abs/2407.18940}.

\bibitem[Artiles et~al.(2026)Artiles, Weiss, Brinkmann, Rahwan, Sch{\"o}lkopf,
  Pal, Larochelle, Goyal, and Rahaman]{artiles2026aliensciencesamplingcoherent}
Artiles, A.~H., Weiss, M., Brinkmann, L., Rahwan, I., Sch{\"o}lkopf, B., Pal,
  C., Larochelle, H., Goyal, A., and Rahaman, N.
\newblock The alien space of science: Sampling coherent but cognitively
  unavailable research directions, 2026.
\newblock URL \url{https://arxiv.org/abs/2603.01092}.

\bibitem[Asai et~al.(2026)Asai, He, Shao, Shi, Singh, Chang, Lo, Soldaini,
  Feldman, D'Arcy, Wadden, Latzke, Sparks, Hwang, Kishore, Tian, Ji, Liu, Tong,
  Wu, Xiong, Zettlemoyer, Neubig, Weld, Downey, Yih, Koh, and
  Hajishirzi]{asai2026openscholar}
Asai, A., He, J., Shao, R., Shi, W., Singh, A., Chang, J.~C., Lo, K., Soldaini,
  L., Feldman, S., D'Arcy, M., Wadden, D., Latzke, M., Sparks, J., Hwang,
  J.~D., Kishore, V., Tian, M., Ji, P., Liu, S., Tong, H., Wu, B., Xiong, Y.,
  Zettlemoyer, L., Neubig, G., Weld, D.~S., Downey, D., Yih, W.-t., Koh, P.~W.,
  and Hajishirzi, H.
\newblock Synthesizing scientific literature with retrieval-augmented language
  models.
\newblock \emph{Nature}, 650:\penalty0 857--863, 2026.
\newblock \doi{10.1038/s41586-025-10072-4}.
\newblock URL \url{https://doi.org/10.1038/s41586-025-10072-4}.

\bibitem[Azerbayev et~al.(2023)Azerbayev, Piotrowski, Schoelkopf, Ayers, Radev,
  and Avigad]{azerbayev2023proofnet}
Azerbayev, Z., Piotrowski, B., Schoelkopf, H., Ayers, E.~W., Radev, D., and
  Avigad, J.
\newblock {ProofNet}: Autoformalizing and formally proving undergraduate-level
  mathematics, 2023.
\newblock URL \url{https://arxiv.org/abs/2302.12433}.

\bibitem[Cohan et~al.(2020)Cohan, Feldman, Beltagy, Downey, and
  Weld]{cohan2020specter}
Cohan, A., Feldman, S., Beltagy, I., Downey, D., and Weld, D.~S.
\newblock {SPECTER}: Document-level representation learning using
  citation-informed transformers.
\newblock In \emph{Proceedings of the 58th Annual Meeting of the Association
  for Computational Linguistics}, 2020.
\newblock URL \url{https://arxiv.org/abs/2004.07180}.

\bibitem[Du et~al.(2025)Du, Xu, Zhu, Wang, and Mao]{du2025deepresearchbench}
Du, M., Xu, B., Zhu, C., Wang, X., and Mao, Z.
\newblock {DeepResearch Bench}: A comprehensive benchmark for deep research
  agents, 2025.
\newblock URL \url{https://arxiv.org/abs/2506.11763}.

\bibitem[Elovic(2023)]{gptresearcher2023}
Elovic, A.
\newblock {GPT Researcher}: Autonomous agent for comprehensive online research.
\newblock Software, 2023.
\newblock URL \url{https://github.com/assafelovic/gpt-researcher}.

\bibitem[{Gemini Team}(2026)]{geminiteam2026gemini31}
{Gemini Team}.
\newblock {Gemini 3.1 Pro}: A smarter model for your most complex tasks.
\newblock Google Blog, 2026.
\newblock URL
  \url{https://blog.google/innovation-and-ai/models-and-research/gemini-models/gemini-3-1-pro/}.

\bibitem[Jaradeh et~al.(2019)Jaradeh, Oelen, Farfar, Prinz, D'Souza,
  Kismih{\'o}k, Stocker, and Auer]{jaradeh2019orkg}
Jaradeh, M.~Y., Oelen, A., Farfar, K.~E., Prinz, M., D'Souza, J., Kismih{\'o}k,
  G., Stocker, M., and Auer, S.
\newblock Open research knowledge graph: Next generation infrastructure for
  semantic scholarly knowledge, 2019.
\newblock URL \url{https://arxiv.org/abs/1901.10816}.

\bibitem[L{\'a}la et~al.(2023)L{\'a}la, O'Donoghue, Shtedritski, Cox,
  Rodriques, and White]{lala2023paperqa}
L{\'a}la, J., O'Donoghue, O., Shtedritski, A., Cox, S., Rodriques, S.~G., and
  White, A.~D.
\newblock {PaperQA}: Retrieval-augmented generative agent for scientific
  research, 2023.
\newblock URL \url{https://arxiv.org/abs/2312.07559}.

\bibitem[{LangChain Team}(2025)]{langchain2025opendeepresearch}
{LangChain Team}.
\newblock Open deep research.
\newblock LangChain Blog and Software, 2025.
\newblock URL \url{https://blog.langchain.com/open-deep-research/}.

\bibitem[Li et~al.(2025)Li, Zeng, Cheng, Ma, and Jia]{li2025reportbench}
Li, M., Zeng, Y., Cheng, Z., Ma, C., and Jia, K.
\newblock {ReportBench}: Evaluating deep research agents via academic survey
  tasks, 2025.
\newblock URL \url{https://arxiv.org/abs/2508.15804}.

\bibitem[Liu et~al.(2023)Liu, Shen, Xin, Liu, Yuan, Wang, Ju, Zheng, Yin, Li,
  Zhang, and Liu]{liu2023fimo}
Liu, C., Shen, J., Xin, H., Liu, Z., Yuan, Y., Wang, H., Ju, W., Zheng, C.,
  Yin, Y., Li, L., Zhang, M., and Liu, Q.
\newblock {FIMO}: A challenge formal dataset for automated theorem proving,
  2023.
\newblock URL \url{https://arxiv.org/abs/2309.04295}.

\bibitem[Lo et~al.(2020)Lo, Wang, Neumann, Kinney, and Weld]{lo2020s2orc}
Lo, K., Wang, L.~L., Neumann, M., Kinney, R., and Weld, D.~S.
\newblock {S2ORC}: The semantic scholar open research corpus, 2020.
\newblock URL \url{https://arxiv.org/abs/1911.02782}.

\bibitem[Lu et~al.(2020)Lu, Dong, and Charlin]{lu2020multixscience}
Lu, Y., Dong, Y., and Charlin, L.
\newblock {Multi-XScience}: A large-scale dataset for extreme multi-document
  summarization of scientific articles.
\newblock In \emph{Proceedings of the 2020 Conference on Empirical Methods in
  Natural Language Processing (EMNLP)}, 2020.
\newblock URL \url{https://arxiv.org/abs/2010.14235}.

\bibitem[McInnes et~al.(2017)McInnes, Healy, and Astels]{mcinnes2017hdbscan}
McInnes, L., Healy, J., and Astels, S.
\newblock {hdbscan}: Hierarchical density based clustering.
\newblock \emph{Journal of Open Source Software}, 2\penalty0 (11):\penalty0
  205, 2017.
\newblock \doi{10.21105/joss.00205}.
\newblock URL \url{https://doi.org/10.21105/joss.00205}.

\bibitem[Pirolli \& Card(1999)Pirolli and Card]{pirolli1999information}
Pirolli, P. and Card, S.~K.
\newblock Information foraging.
\newblock \emph{Psychological Review}, 106\penalty0 (4):\penalty0 643--675,
  1999.
\newblock \doi{10.1037/0033-295X.106.4.643}.

\bibitem[Priem et~al.(2022)Priem, Piwowar, and Orr]{priem2022openalex}
Priem, J., Piwowar, H., and Orr, R.
\newblock Openalex: A fully-open index of scholarly works, authors, venues,
  institutions, and concepts, 2022.
\newblock URL \url{https://arxiv.org/abs/2205.01833}.

\bibitem[Sahu et~al.(2025)Sahu, Larochelle, Charlin, and
  Pal]{sahu2025reviewertoo}
Sahu, G., Larochelle, H., Charlin, L., and Pal, C.
\newblock {ReviewerToo}: Should ai join the program committee? a look at the
  future of peer review, 2025.
\newblock URL \url{https://arxiv.org/abs/2510.08867}.

\bibitem[Shang et~al.(2025)Shang, Wan, Peng, Wu, Chen, Yan, and
  Zhang]{shang2025stepfun}
Shang, S., Wan, R., Peng, Y., Wu, Y., Chen, X.-h., Yan, J., and Zhang, X.
\newblock {StepFun-Prover} preview: Let's think and verify step by step, 2025.
\newblock URL \url{https://arxiv.org/abs/2507.20199}.

\bibitem[Shao et~al.(2024)Shao, Jiang, Kanell, Xu, Khattab, and
  Lam]{shao2024storm}
Shao, Y., Jiang, Y., Kanell, T.~A., Xu, P., Khattab, O., and Lam, M.~S.
\newblock Assisting in writing {Wikipedia}-like articles from scratch with
  large language models.
\newblock In \emph{Proceedings of the 2024 Conference of the North American
  Chapter of the Association for Computational Linguistics: Human Language
  Technologies}, 2024.
\newblock URL \url{https://arxiv.org/abs/2402.14207}.

\bibitem[Shin et~al.(2025)Shin, Tang, Lee, Kim, Lim, Cho, Hong, Lee, and
  Kim]{shin2025blindspots}
Shin, H., Tang, J., Lee, Y., Kim, N., Lim, H., Cho, J.~Y., Hong, H., Lee, M.,
  and Kim, J.
\newblock Mind the blind spots: A focus-level evaluation framework for {LLM}
  reviews, 2025.
\newblock URL \url{https://arxiv.org/abs/2502.17086}.

\bibitem[Thilakaratne et~al.(2019)Thilakaratne, Falkner, and
  Atapattu]{thilakaratne2019systematic}
Thilakaratne, M., Falkner, K., and Atapattu, T.
\newblock A systematic review on literature-based discovery workflow.
\newblock \emph{PeerJ Computer Science}, 5:\penalty0 e235, 2019.
\newblock \doi{10.7717/peerj-cs.235}.
\newblock URL \url{https://doi.org/10.7717/peerj-cs.235}.

\bibitem[Wadden et~al.(2020)Wadden, Lin, Lo, Wang, van Zuylen, Cohan, and
  Hajishirzi]{wadden2020fact}
Wadden, D., Lin, S., Lo, K., Wang, L.~L., van Zuylen, M., Cohan, A., and
  Hajishirzi, H.
\newblock Fact or fiction: Verifying scientific claims.
\newblock In \emph{Proceedings of the 2020 Conference on Empirical Methods in
  Natural Language Processing}, 2020.
\newblock URL \url{https://arxiv.org/abs/2004.14974}.

\bibitem[Ye et~al.(2025)Ye, Yan, He, Kasriel, Yang, and Song]{ye2025verina}
Ye, Z., Yan, Z., He, J., Kasriel, T., Yang, K., and Song, D.
\newblock {VERINA}: Benchmarking verifiable code generation, 2025.
\newblock URL \url{https://arxiv.org/abs/2505.23135}.

\bibitem[Zhao et~al.(2024)Zhao, Lv, Xu, Wei, Wang, Dang, Liu, and
  Chen]{zhao2024detrs}
Zhao, Y., Lv, W., Xu, S., Wei, J., Wang, G., Dang, Q., Liu, Y., and Chen, J.
\newblock {DETRs} beat {YOLOs} on real-time object detection.
\newblock In \emph{Proceedings of the IEEE/CVF Conference on Computer Vision
  and Pattern Recognition}, 2024.
\newblock URL \url{https://arxiv.org/abs/2304.08069}.

\end{thebibliography}
\bibliographystyle{icml2026}

\onecolumn
\appendix

\section{Per-Claim Audit}
\label{app:audit}

Table~\ref{tab:audit-full} summarizes the detailed grounding audit. Each row pairs an extracted claim with the \texttt{/md} page (and section) where its support is located.

\begin{table}[h]
\centering
\small
\begin{tabular}{@{}p{0.28\textwidth}p{0.34\textwidth}p{0.10\textwidth}p{0.18\textwidth}@{}}
\toprule
Claim & Evidence checked & Status & Notes \\
\midrule
Autoformalization is hard because informal mathematics leaves assumptions implicit while proof assistants require explicit formal structure. & Autoformalization-gap research direction page; ProofNet paper page (typechecking, implicit hypotheses, retrieval). & Supported & Concrete example: orthogonal complement presupposes inner product space. \\
\addlinespace
Compiler feedback turns formal theorem proving from one-shot generation into iterative correction. & Iterative-compiler-feedback research direction page; FIMO; StepFun-Prover; VERINA. & Supported & FIMO and StepFun document the feedback loop; VERINA documents costs. \\
\addlinespace
Iterative feedback alone does not solve deep logical gaps. & VERINA; iterative-compiler-feedback research direction page (Semantic Wall section). & Supported (limitation) & 64-round VERINA refinement plateaus near 20\% ProofGen. \\
\addlinespace
Token-level alignment of informal mathematics to formal compiler states is a validated result. & Token-level alignment research proposal page. & Not supported (research-proposal-level) & The research proposal page describes a planned method, not a result. \\
\addlinespace
The run is perfectly clean and contains only theorem-proving papers. & Research direction context and related-paper tabs. & False (useful failure case) & The run includes some adjacent or noisy papers from broader verifiable-reasoning neighborhoods. \\
\bottomrule
\end{tabular}
\caption{Full per-claim audit for the autoformalization run, with source, support status, and audit note.}
\label{tab:audit-full}
\end{table}

\section{Map Construction Pipeline}
\label{app:construction}

Lacuna's hosted map is built by a pipeline running on production infrastructure. The pipeline starts from scholarly records, paper files, author metadata, and venue metadata. LLM stages then create core-idea paper summaries, paper summaries with figures, concept elements, research directions, and research proposals, while preserving the database links needed to follow generated content back to the records that produced it. The production map used the Gemini-family models named in Section~\ref{sec:map}. Table~\ref{tab:pipeline-stages} summarizes the full construction pipeline; Figure~\ref{fig:pipeline} in the main text shows the paper-to-direction generation subpipeline.

\begin{table}[H]
\caption{Major construction stages, durable database outputs, and their role in the hosted Lacuna map.}
\label{tab:pipeline-stages}
\centering
\small
\begin{tabular}{p{0.18\textwidth}p{0.30\textwidth}p{0.42\textwidth}}
\toprule
Stage group & Primary durable outputs & Role in the map \\
\midrule
Catalog harvest and reconciliation & Paper, author, venue, identifier, and authorship records; duplicate and alias resolutions & Establishes stable paper, author, and venue records before any generated synthesis is attached. \\
\addlinespace
Source acquisition & \texttt{blob\_metadata} file identities and \texttt{blob\_location} replicas for PDFs or URLs & Separates the identity of a paper file from where it can be fetched, so the system can attach new replicas without changing the paper object. \\
\addlinespace
Paper text processing & Core-idea paper summaries, paper tags, and explanation records & Produces the text used for concept extraction and for readable paper pages. \\
\addlinespace
Figure-rich paper summary processing & Extracted figures, selected figures, captions, and paper summaries with figures & Produces compact paper summaries that combine text and selected visual evidence. \\
\addlinespace
Concept extraction & Concept elements, embeddings, paper assignments, and provenance snippets & Converts paper-level observations into one- or two-sentence units that can be clustered, searched, and linked back to papers. \\
\addlinespace
Research direction clustering & \texttt{taxonomy\_node} rows and \texttt{concept\_element\_assignment} links & Groups recurring technical problems, methods, and opportunity areas into research direction pages with representative papers and neighboring research directions. \\
\addlinespace
Research proposal generation & Alien Science samples and research proposal pages & Uses paper summaries with figures, concept elements, and research direction context to generate candidate hypotheses. \\
\addlinespace
Serving indexes and exports & Search indexes, lookup indexes, relation tables, HTML pages, \texttt{/md} variants, and MCP access & Makes the map usable through search, tabs, hover previews, pagination, typed links, markdown pages, and tool calls. \\
\bottomrule
\end{tabular}
\end{table}

\textbf{Authorship-anchored bibliography expansion.}
Resolved author identities are built around OpenReview-maintained records. For each OpenReview-profiled author \(a\), the catalog pipeline first harvests the author's OpenReview papers, including records available only through legacy OpenReview APIs, and stores their positional author lists. The pipeline canonicalizes aliases and aligns raw author identifiers to author names before using them as person-level links, so older records with aliases, emails, or other legacy identifiers do not fragment a single author into multiple resolved identities.

The pipeline then constructs a trusted co-author neighborhood \(C_a\) from OpenReview papers only. Each edge records a normalized co-author name, an optional resolved co-author identifier, and the number of shared OpenReview papers. This OpenReview-only neighborhood is used as an authorship signal for external bibliography expansion. External identifiers declared by OpenReview records are followed into DBLP and OpenAlex to obtain candidate papers \(E_a\). For each candidate paper, the pipeline reads the source's author list, normalizes names, and applies the following rule: retain the paper for \(a\) if it reconciles to an existing OpenReview paper for \(a\), or if at least one candidate co-author matches \(C_a\); otherwise the candidate is left unattached to the resolved author identity unless later reconciliation provides stronger evidence. Candidates from authors with too little OpenReview co-author evidence are deferred, not treated as high-confidence rejections.

After this filtering step, paper reconciliation merges duplicate source records using identifiers, normalized titles, years, and author overlap, preferring canonical records that preserve OpenReview provenance when available. The resulting bibliography for an author is therefore the union of OpenReview papers and externally harvested papers whose authorship is supported by the OpenReview co-author graph. This keeps external-source disambiguation errors from propagating into generated summaries, research-direction memberships, and proposal context.

\textbf{Records.} Lacuna treats source records, source files, metadata entities, and generated content as separate records. A paper page is keyed by a canonical artifact identifier; its files are keyed by content-hash-like blob metadata; its core-idea paper summary, paper summary with figures, extracted figures, concept elements, research direction memberships, and research proposal links are stored as derived records. This separation matters in practice: a paper can remain searchable if a PDF URL dies, a new open-access replica can be attached without changing the paper identity, and generated synthesis can be regenerated while preserving links to the underlying records.

\textbf{Generated Content and Links.} The pipeline does not treat model-written text as free-floating prose. Concept elements are assigned back to papers; research direction articles are built from clusters of concepts and representative papers; research proposal pages list the research directions and papers they use as context. The serving layer preserves these relationships as visible tabs, evidence sections, and markdown links. This is why the claim audit in Section~\ref{sec:grounding} can check generated claims against specific page sections.

\textbf{Pipeline organization.} Pipeline stages are numbered, rerunnable, and idempotent at the stage level: harvest and reconciliation stages build the normalized catalog; PDF/blob stages acquire and verify paper files; batch LLM stages write core-idea paper summaries, concept elements, research direction articles, research proposals, and paper summaries with figures; embedding and clustering stages build the research direction graph; serving stages export lookup tables and search indexes. Large batch jobs run on production GCE infrastructure and write durable counters and manifests.

\section{ScholarQA-CS-ML Evaluation Details}
\label{app:scholarqa}

ScholarQA-CS (called Scholar-CS in the OpenScholar paper) is the computer-science subset of ScholarQABench~\citep{asai2026openscholar}. It contains 100 literature-review questions written by expert annotators holding PhDs; each question has detailed answer rubrics built from literature evidence and supporting quotes. The public test configuration we used contains 100 unique prompts.

To define the ML/AI slice, Codex GPT-5.5 with xhigh reasoning effort was instructed to act as a classifier and select prompts about machine learning, artificial intelligence, NLP, or robot learning. This produced 42 questions, or 42\% of ScholarQA-CS. We evaluated all 42 selected prompts as-is, without tuning or filtering by Lacuna or OpenScholar performance.

Lacuna-GPT-4o used abstract-only Elasticsearch evidence from Lacuna, a facet-conditioned retrieval plan, top-15 paper evidence, concept evidence packets, and compact cited evidence appended to the answer with original citation IDs preserved. It reached an average rubric score of 0.694 over the 42 selected questions. On the same 42 questions, the released OpenScholar-GPT-4o answer file scored 0.672, while the same OpenScholar answers with appended references stripped scored 0.589. Lacuna-GPT-4o without the appended cited-evidence block scored 0.572. We report both answer-only and answer-plus-evidence formats because the released OpenScholar file itself includes appended reference material; the comparison is therefore both a synthesis check and a check that Lacuna can serve useful source context to the answerer. The component breakdown in Table~\ref{tab:scholarqa} shows that Lacuna-GPT-4o improves the content-side rubric metrics while OpenScholar-GPT-4o retains higher structural citation coverage.

\section{Additional Benchmark Details}
\label{app:additional-benchmarks}

This appendix records the benchmark-scope names used in the paper. LitSearch keeps the original benchmark name because the benchmark is already ML/NLP-centered and we now evaluate its full query split. Multi-XScience-CS/ML is a CS/ML-filtered pilot subset of Multi-XScience. ScholarQA-CS-ML is the ML/AI slice of the ScholarQA computer-science subset. ReportBench-ML is the 25-task machine-learning subset of ReportBench.

\begin{table}[t]
\caption{Benchmark protocols for the literature search, synthesis, and report-generation results.}
\label{tab:benchmark-protocols}
\centering
\small
\begin{tabular}{@{}p{0.15\textwidth}p{0.24\textwidth}p{0.15\textwidth}p{0.18\textwidth}p{0.18\textwidth}@{}}
\toprule
Benchmark & Task & Split & Lacuna input & Primary scoring \\
\midrule
LitSearch & Retrieve target papers from natural-language paper descriptions. & Full 597-query split. & Lacuna production paper search. & MRR and recall at 10. \\
Multi-XScience-CS/ML & Write related-work synthesis from cited-paper context. & 30 CS/ML-filtered test examples. & Lacuna evidence for available cited papers. & LLM-judged score out of 5. \\
ScholarQA-CS-ML & Answer ML/AI literature-review questions with grounding. & 42 ML/AI prompts from ScholarQA-CS. & Lacuna paper, concept, and cited-evidence packets. & ScholarQABench rubric average. \\
ReportBench-ML & Write survey-style deep-research reports. & 25 core-ML survey tasks. & Lacuna Deep Research reports. & Citation overlap and RACE quality. \\
\bottomrule
\end{tabular}
\end{table}

\subsection{LitSearch}
\label{app:litsearch}

LitSearch is a retrieval-only benchmark in this evaluation: no answer generator or judge model is called. We loaded the HuggingFace \texttt{princeton-nlp/LitSearch} query split (597 queries) and \texttt{corpus\_clean} split. The full split contains 574 unique gold \texttt{corpusid} papers. For Lacuna matching, 567 of those gold papers have titles in \texttt{corpus\_clean}; a Lacuna gold-resolution cache resolves 544 unique gold papers, giving at least one resolved gold paper for 575 of 597 queries. We score all 597 queries and count a hit if any returned paper matches a gold paper by resolved Lacuna artifact id or normalized title.

The reported Lacuna row uses the deployed production default paper search with \texttt{type=paper}, \texttt{limit=10}, \texttt{sort=relevance}, \texttt{ranking\_profile=default}, and \texttt{strict=1}. The OpenScholar v3 row uses the released OpenScholar-v3 retrieval artifacts: the OSDS datastore, \texttt{akariasai/pes2o\_contriever} top-100 candidates, and \texttt{OpenScholar/OpenScholar\_Reranker} final ranking. Because OpenScholar retrieves chunks, the reported row deduplicates reranked chunks by paper id before computing top-10 metrics.

\begin{table}[h]
\caption{LitSearch full-benchmark retrieval details.}
\label{tab:litsearch-details}
\centering
\small
\begin{tabular}{@{}lrrrr@{}}
\toprule
System & R@1 & R@5 & R@10 & MRR \\
\midrule
Lacuna default strict & \textbf{0.278} & \textbf{0.466} & \textbf{0.538} & \textbf{0.3585} \\
OpenScholar v3 & 0.256 & 0.377 & 0.424 & 0.3081 \\
\bottomrule
\end{tabular}
\end{table}

These values are the paper-level retrieval metrics for the full LitSearch query split.

\subsection{Multi-XScience-CS/ML}
\label{app:multixscience}

Multi-XScience-CS/ML uses the public Multi-XScience test split and keeps 30 examples selected by scanning the split for CS/ML-ish target papers using ML/AI keywords and OpenAlex concepts. Multi-XScience exposes cited-paper MAG ids and abstracts, not cited titles, so cited papers are first resolved through OpenAlex and then matched to Lacuna by title search with normalized title similarity at least 0.92. Lacuna evidence consists of generated paper-page summaries and concept elements for matched cited papers, capped at six cited papers per example. The selected examples average 7.90 original cited-paper abstracts and 3.77 Lacuna matched cited papers used in generation, for a mean cited-reference coverage fraction of 0.5586.

Generation and judging use \texttt{models/gemini-3.1-pro-preview}. The main table reports the retained no-date-limit Lacuna row (overall 4.167) to match the benchmark summary. A fixed-selection rerun against the current server reached overall 4.200, while the same-run original-abstract ceiling was 4.333. The OpenScholar-v3 comparison uses the released OpenScholar-v3 retrieval artifacts, including OSDS top-100 \texttt{pes2o\_contriever} candidates and \texttt{OpenScholar/OpenScholar\_Reranker} final ranking, with the same 30 examples, generation prompt, judge prompt, and judge model.

\begin{table}[h]
\caption{Multi-XScience-CS/ML related-work synthesis details. Scores are 1--5 Gemini judge means.}
\label{tab:multixscience-details}
\centering
\small
\begin{tabular}{@{}lrrrrr@{}}
\toprule
Condition & Rel. & Cov. & Spec. & Supp. & Overall \\
\midrule
Lacuna, reported row & -- & -- & -- & -- & 4.167 \\
Lacuna, current server & 4.850 & 4.133 & 4.617 & 4.317 & 4.200 \\
Original cited abstracts & 4.900 & 4.567 & 4.500 & 4.100 & 4.333 \\
Target abstract only & 2.700 & 1.300 & 1.533 & 4.133 & 2.067 \\
OpenScholar v3 & 4.367 & 3.533 & 4.000 & 3.200 & 3.467 \\
\bottomrule
\end{tabular}
\end{table}

The main limitation is citation coverage: the Lacuna condition uses only matched cited papers, while the original-abstract baseline uses all cited abstracts provided by the benchmark. In the cutoff-matched match analysis, 185 OpenAlex-title searches were attempted for cited papers, 134 accepted Lacuna matches were found, 51 cited papers had no accepted match, and 113 matched paper slots were used after the per-example cap. Missing papers cluster in older or adjacent venues such as journals, LNCS, IEEE proceedings, and repositories rather than core modern ML venues.

{\scriptsize
\begin{longtable}{p{0.12\textwidth}p{0.12\textwidth}p{0.66\textwidth}}
\caption{Multi-XScience-CS/ML selected target papers.}
\label{tab:multixscience-selected}\\
\toprule
Example ID & arXiv ID & Target title \\
\midrule
\endfirsthead
\toprule
Example ID & arXiv ID & Target title \\
\midrule
\endhead
test\_20 & 1908.11078 & Document Hashing with Mixture-Prior Generative Models \\
test\_21 & 1908.11314 & Variational Denoising Network: Toward Blind Noise Modeling and Removal \\
test\_23 & 1908.11057 & A Deep Neural Information Fusion Architecture for Textual Network Embeddings \\
test\_24 & 1908.10797 & Image Captioning with Sparse Recurrent Neural Network \\
test\_29 & 1908.10084 & Sentence-BERT: Sentence Embeddings using Siamese BERT-Networks \\
test\_30 & 1908.10155 & Mobile Video Action Recognition \\
test\_36 & 1908.09506 & Constraint learning for control tasks with limited duration barrier functions \\
test\_44 & 1908.08326 & Revisit Semantic Representation and Tree Search for Similar Question Retrieval. \\
test\_49 & 1908.06893 & Automated email Generation for Targeted Attacks using Natural Language \\
test\_54 & 1908.05908 & BERT-Based Multi-Head Selection for Joint Entity-Relation Extraction \\
test\_60 & 1908.05498 & A Single-Shot Arbitrarily-Shaped Text Detector based on Context Attended Multi-Task Learning \\
test\_75 & 1908.03477 & Fine-Grained Action Retrieval Through Multiple Parts-of-Speech Embeddings \\
test\_83 & 1908.02711 & I Bet You Are Wrong: Gambling Adversarial Networks for Structured Semantic Segmentation \\
test\_85 & 1908.02256 & BlurNet: Defense by Filtering the Feature Maps \\
test\_88 & 1908.01536 & Discriminating Spatial and Temporal Relevance in Deep Taylor Decompositions for Explainable Activity Recognition \\
test\_93 & 1908.00948 & High-Level Control of Drum Track Generation Using Learned Patterns of Rhythmic Interaction \\
test\_94 & 1908.00355 & Continual Learning via Online Leverage Score Sampling \\
test\_96 & 1908.00222 & Structured3D: A Large Photo-realistic Dataset for Structured 3D Modeling \\
test\_102 & 1907.13315 & Self-training with progressive augmentation for unsupervised cross-domain person re-identification \\
test\_104 & 1907.12679 & Machine Translation Evaluation with BERT Regressor \\
test\_112 & 1907.12006 & ROAM: Recurrently Optimizing Tracking Model \\
test\_113 & 1907.11840 & Learning Instance-wise Sparsity for Accelerating Deep Models \\
test\_119 & 1907.11397 & Improving Generalization via Attribute Selection on Out-of-the-box Data \\
test\_123 & 1907.10202 & Title unavailable in source record \\
test\_124 & 1907.10107 & Lifelong GAN: Continual Learning for Conditional Image Generation \\
test\_128 & 1907.10156 & DR Loss: Improving Object Detection by Distributional Ranking \\
test\_134 & 1907.08195 & Temporally Coherent General Dynamic Scene Reconstruction \\
test\_144 & 1901.11467 & Towards Controlled Transformation of Sentiment in Sentences \\
test\_148 & 1901.11409 & Semantic Redundancies in Image-Classification Datasets: The 10\% You Don't Need \\
test\_153 & 1901.11150 & Title unavailable in source record \\
\bottomrule
\end{longtable}
}

\subsection{ReportBench-ML}
\label{app:reportbench}

ReportBench-ML uses 25 core-ML survey tasks derived from peer-reviewed arXiv survey papers. ReportBench scores a generated report by comparing the papers it cites with the expert survey's reference list. The citation-overlap row uses title-only matching without URL matching: cited titles are extracted with GPT-4o for every system and matched to survey references by normalized exact title. The RACE row uses the DeepResearch Bench report-quality rubric with GPT-5.5 as judge; because RACE is comparative, the ranking and broad gaps are more meaningful than the last decimal.

The Lacuna row is v11 of the production Lacuna Deep Research implementation described in Section~\ref{sec:deep-research-framework}: fixed seed-direction search by extracted topic, high-effort lens workers, constraint extraction, requirement-aware synthesis, and audit/revision. Citation metrics are computed on all 25 non-empty v11 reports after retrying one length-cap failure. The GPT-Researcher row uses the package defaults for research-report generation; those defaults do not pin a single LLM and retriever environment. STORM uses Exa retrieval, \texttt{gpt-4o-mini} for conversation/question asking, and \texttt{gpt-4o} for outline, article generation, and polish. LangChain Open Deep Research intentionally uses Gemini 3.5 Flash for research, compression, and final report generation.

Expert surveys cite roughly 140 papers on average, while the evaluated deep-research agents cite about 6--16 papers per survey. This makes recall structurally small for every system and makes precision, F1, and total expert-reference hits more informative than recall alone.

{\scriptsize
\begin{longtable}{p{0.12\textwidth}p{0.78\textwidth}}
\caption{ReportBench-ML task IDs and survey titles.}
\label{tab:reportbench-ml-tasks}\\
\toprule
arXiv ID & Survey title \\
\midrule
\endfirsthead
\toprule
arXiv ID & Survey title \\
\midrule
\endhead
2407.03993 & A Survey on Natural Language Counterfactual Generation \\
2204.10365 & Towards an Enhanced Understanding of Bias in Pre-trained Neural Language Models: A Survey with Special Emphasis on Affective Bias \\
2206.04149 & A Survey of Graph-based Deep Learning for Anomaly Detection in Distributed Systems \\
2109.12843 & A Survey of Graph Neural Networks for Recommender Systems: Challenges, Methods, and Directions \\
2302.00058 & Graph Anomaly Detection in Time Series: A Survey \\
2310.11011 & From Identifiable Causal Representations to Controllable Counterfactual Generation: A Survey on Causal Generative Modeling \\
2401.16386 & Continual Learning with Pre-Trained Models: A Survey \\
2007.08199 & Learning from Noisy Labels with Deep Neural Networks: A Survey \\
2003.00653 & Adversarial Attacks and Defenses on Graphs: A Review, A Tool and Empirical Studies \\
2404.00929 & A Survey on Multilingual Large Language Models: Corpora, Alignment, and Bias \\
2304.11534 & Graph Neural Networks for Text Classification: A Survey \\
2406.07494 & CADS: A Systematic Literature Review on the Challenges of Abstractive Dialogue Summarization \\
2103.07853 & Membership Inference Attacks on Machine Learning: A Survey \\
2209.00099 & Efficient Methods for Natural Language Processing: A Survey \\
2212.04634 & Open-world Story Generation with Structured Knowledge Enhancement: A Comprehensive Survey \\
2305.08493 & Creative Data Generation: A Review Focusing on Text and Poetry \\
2306.05817 & How Can Recommender Systems Benefit from Large Language Models: A Survey \\
2407.15186 & A Survey on Employing Large Language Models for Text-to-SQL Tasks \\
2207.07483 & A Systematic Review and Replicability Study of BERT4Rec for Sequential Recommendation \\
2011.13534 & A Survey of Deep Learning Approaches for OCR and Document Understanding \\
2111.00358 & A Survey on the Robustness of Feature Importance and Counterfactual Explanations \\
2202.06481 & A Survey on Machine Learning Approaches for Modelling Intuitive Physics \\
2402.05617 & Deep Learning-based Computational Job Market Analysis: A Survey on Skill Extraction and Classification from Job Postings \\
2408.00516 & Low-Power Vibration-Based Predictive Maintenance for Industry 4.0 using Neural Networks: A Survey \\
2111.04006 & A Review of Location Encoding for GeoAI: Methods and Applications \\
\bottomrule
\end{longtable}
}

{\scriptsize
\begin{longtable}{p{0.33\textwidth}p{0.59\textwidth}}
\caption{ScholarQA-CS-ML question IDs and prompts selected by ML/AI topic classification.}
\label{tab:scholarqa-ml-questions}\\
\toprule
Question ID & Prompt \\
\midrule
\endfirsthead
\toprule
Question ID & Prompt \\
\midrule
\endhead
00bdd80debc8549198001289188c6bea & How can I use an hybridization of ontology and machine learning for text summarization ? \\
018bff91263ab3be5f8ad5bade76b030 & In robotics, what are the leading methods for learning terrain traversibility costs automatically from robot experience? \\
0348920a58979e759af9081a6225ee0d & What are the advantages and limitations of applying bias mitigation algorithms during pre-processing, training and inference stages of a model? \\
0919a8528cd20166163de3fdcb089efa & Have large langauge models been applied to the schema matching problem in databases and are they effective? \\
11e71107ddfdc824b8b87d4f5a2ef843 & What data preprocessing steps are most important for point cloud datasets before performing surface reconstruction? \\
15dec998cf77887f870ebf9a55bb7e89 & How effective are language models at automatically generating textual descriptions of scientific concepts? \\
1f384c4d3942c46b676b5bfc66447192 & How good are LLMs at solving traditional tabular ML datasets using ICL? \\
2bb40aa93ac3a6a673c839bd660718ac & Have specialized approaches been developed for providing LLM assistance when people author SQL queries? \\
2dc0ff181a621680dde0a48e0d63f0d9 & How has the citation graph been used to improve neural language models for scientific papers? \\
323e85c9052082358fc0c045fe20a537 & How do different fairness metrics correlate with each other across various datasets and model architectures? \\
4534bd4b99ea2bfd1efd8c656e9264c7 & What datasets and methods are used to pre-train models on table specific tasks? \\
5acd0e1d36af3c52c3159b4b230bcc2f & What interfaces have researchers developed to help people perform behavioral evaluation of ML models, and how do they accelerate annotation efforts? \\
6f526f72804ce3eb59feb7046f319ccc & What are good practices for detecting AI-generated texts in situations where false positives are extremely expensive? \\
803bc7891917f823a52948ebee89cf9d & How does in-context learning for LLMs differ from traditional machine learning model training? \\
928646729f5d824dd9a52e9ddff70e58 & Describe what is known about overfitting in in-context learning. \\
948b6cb986a5d7732722975dbed9d420 & What are leading approaches for evaluating complex scientific question answering systems in NLP? \\
a8ba07610b6d77890e50144bfd4d4168 & What are some common UI designs for sense-making, information organization, and AI writing tools? \\
ab3651c422a54f40aa4bac3fa630c13e & How do large language models like ChatGPT impact the diversity of published scientific ideas? \\
baee287ff68f0fd60dcbd0d8b9b741b8 & Apart from preventing overfitting, are there any side effects (desirable or otherwise) of applying dropout in deep neural networks? \\
bb7198e650267504d67b14b6e4c67c7c & what are the most important open challenges in using neural networks in combination with PDE solvers for fluid simulation? \\
be5c0337461175e55f2a8fa9bcce5732 & What are the leading approaches for anomoly detection in process mining? \\
ce433b751f38f7d0173095c2faa2f75a & What are advantages and disadvantages of top methods for picking the right number of topics in topic modeling? \\
e2d0eb391fbf674c070c74c387ee6248 & During pre-training, why is the transformer embedding layer initialized randomly, rather than with pre-trained embeddings from existing models? \\
e894bc20daf0522da9c576ae27b257de & What are the main challenges in adapting transformer-based models for tabular data representation? \\
fb607bc177d2efb926cb3dff15668861 & How can question generation be used to mitigate hallucination in LLMs? \\
fc927b39177cd2aad8a8cbcef75ab62c & What are the leading thoughts about how to teach computer science to college students in the age of generative AI? \\
0ed1770483ec64633a580366026dd16e & What are some of the challenges associated with creating a training dataset for question answering in scientific domains? Describe some recent methods that try to overcome these challenges. \\
34b00939c1190990c1c9be590cb07476 & How has literature discussed AI as a design material? \\
3d8c315aed4cc104f2ad61f4deeda9c1 & What are the leading techniques for person-following robot navigation (which must track the person and potentially re-identify if occluded)? \\
5480ba91951fc42e9beb989eea40360d & How does the addition of XAI techniques such as SHAP or LIME impact model interpretability in complex machine learning models ? \\
6453c86c72949978fb8db5405b48c923 & is there any evidence that large language models can be effectively applied to robot planning tasks? \\
64ef9b9e4c220dd1a2f6115b2a9e242a & What are the most important open challenges in Federated Learning? \\
6e7e3524d565599b3064ae05375956f7 & what are the best recent techniques for text watermarking, what kinds of transformations are they robust to, and what transformations or attacks are they not robust against \\
7670af632f0932d5ed20c34e1c9f01d2 & What interfaces have researchers developed for helping people optimize LLMs for a task, and what are the biggest remaining user problems? \\
983e73defc06e6794a856330905dc787 & Does active learning work well when fine-tuning large language models? \\
a5d23eb3a2555db0a82f6b64fed85baa & What are some tasks where fine tuning smaller models is beneficial over using LLMs? \\
aab38dd1282ff3387cc8bf9bdf13b3aa & What are the leading approaches to automatic scientific paper review generation and what are their strengths and weaknesses? \\
b069a1248503c4caa98dab0014d1a55a & In recommendation systems, how are new methods that optimize diversity typically evaluated? \\
e03c49cccc971cf3ae67556554b4666b & Are there papers that use different formats of Q\&A with the user to clarify intent and compose more complicated prompts to LLM? \\
e09a30179e0b587d57edb17317ae3288 & How is artificial intelligence being utilized to enhance the diagnosis and treatment of sleep apnea? \\
e61be146ef53f1a5483c668fc4f6390c & What principles of interactive machine teaching can be applied to interactively curating social media feeds? \\
eef3ee38231d9fa5ffabbec75d1c5b50 & What are good benchmarks and evaluation strategies for comparing obstacle avoidance strategies in mobile robots? \\
\bottomrule
\end{longtable}
}

\end{document}